\documentclass[conference]{IEEEtran}
\IEEEoverridecommandlockouts

\usepackage[noend]{algpseudocode}
\usepackage{algorithm}
\usepackage{amsmath}
\usepackage{cite}
\usepackage{graphicx}
\usepackage{makecell}
\usepackage[braket, qm]{qcircuit}
\usepackage{subfigure}
\usepackage{url}
\usepackage[table]{xcolor}
\usepackage{multirow}
\usepackage[switch]{lineno} 

\definecolor{Celadon}{RGB}{175, 225, 175}
\def\BibTeX{{\rm B\kern-.05em{\sc i\kern-.025em b}\kern-.08em
    T\kern-.1667em\lower.7ex\hbox{E}\kern-.125emX}}

\begin{document}
\title{Boosting the Efficiency of Quantum Divider through Effective Design Space Exploration
}

\author{
\IEEEauthorblockN{Siyi Wang, Eugene Lim and Anupam Chattopadhyay}
\IEEEauthorblockA{School of Computer Science and Engineering\\
Nanyang Technological University, Singapore\\
\textit{siyi002@e.ntu.edu.sg}}
}

\maketitle

\begin{abstract}
Rapid progress in the design of scalable, robust quantum computing necessitates efficient quantum circuit implementation for algorithms with practical relevance. For several algorithms, arithmetic kernels, in particular, division plays an important role.
In this manuscript, we focus on enhancing the performance of quantum slow dividers by exploring the design choices of its sub-blocks, such as, adders. Through comprehensive design space exploration of state-of-the-art quantum addition building blocks, our work have resulted in an impressive achievement: a reduction in Toffoli Depth of up to $94.06\%$, accompanied by substantial reductions in both Toffoli and Qubit Count of up to $91.98\%$ and $99.37\%$, respectively.
This paper offers crucial perspectives on efficient design of quantum dividers, and emphasizes the importance of adopting a systematic design space exploration approach.

\end{abstract}


\section{Introduction}

As a fundamental arithmetic operation in both classical and quantum computing, the development of division is crucial in important quantum algorithms such as computing shifted quadratic character problems \cite{Hallgren_2007}, principal ideal problems \cite{VanDam_2001} and hidden shift problems \cite{VanDam_2002}. 

In the past, significant effort was spent to design and optimize division for classical computing \cite{Thapliyal_2017} \cite{Thapliyal_2021} \cite{Gayathri_2021} \cite{Gayathri_2022}. In quantum computing, this problem is relatively unexplored. As we will show in this work, there is significant room for improving the performance. Moreover, we emphasize the need for a systematic design space exploration for building large scale designs. To aid future research and benchmarking, we present detailed design steps apart from offering open-sourced circuit implementation\footnote{The relevant code will be available as a public repository (\url{https://github.com/Siyi-06/Quantum_Non_Restoring_Divider}).}. 

Design Space Exploration (DSE) is a standard practice for designing large circuits in classical computing~\cite{Ma_2019}. However, it is not yet applied to the quantum circuit designs of quantum division. We begin by decomposing non-restoring division into various sub-modules, prominently consisting of adders. Since there are many quantum adders offering diverse trade-offs in logical depth and size (gate count, qubit count), the natural question arises about the selection of a specific quantum adder. 
We also ensure that the proposed circuit is based on the Clifford+T gate set. This is crucial in permitting fault tolerant and scalable quantum computation. Although many different metrics are available to measure the performance of a quantum circuit, we evaluate Toffoli Depth (TD), Toffoli Count (TC) and Qubit Count (QC) as they form a commonly accepted basis for comparison. They can also be easily expressed in terms of T-Depth/T-Count \cite{T_depth_one_2013} if one considers the prevalent Clifford $+$ T gate library.

This paper makes the following substantial contributions.
\begin{itemize}
\item Comprehensive application of DSE methodology during quantum division circuit implementation
\item Achievement of unprecedented performance in quantum division circuits through careful consideration of all relevant metrics
\end{itemize}


\section{Related Work}\label{sec: Prev}
In this section, we systematically review the existing literature on classical and quantum division designs.

In the classical world, division algorithms can be categorized as slow and fast methods \cite{Gayathri_2022}. 
Slow division methods, including restoring and non-restoring methods, iteratively calculate partial remainders through multiple iterations involving addition and subtraction operations, eventually arriving at the final quotient.
While significantly reducing time costs compared to slow division, fast division methods require complex conditional statements, resource-intensive arithmetic operations, or higher radix.

In the quantum world, researchers have explored various efficient division circuit designs following the classifications in classical division.
For quantum fast division, Mathias et al. \cite{Mathias_2017} demonstrated the synthesis of a quantum Newton-Raphson divider to show that classical logic synthesis algorithms can be used to convert well established hardware description languages into quantum circuits. 
In 2021, Gayathri et al. \cite{Gayathri_2021} proposed a novel Goldschmidt divider. 
Afterwards, Gayathri et al. \cite{Gayathri_2022} introduced a quantum Newton-Raphson division circuit, significantly reducing TD, TC and QC from the previous designs. 
For quantum slow division, Khosropour et al. \cite{Khosropour_2011} presented an early high level restoring division design. 
After that, Thapliyal et al. \cite{Thapliyal_2017} \cite{Thapliyal_2021} proposed specific quantum circuits for restoring and non-restoring division.
However, their circuits \cite{Thapliyal_2021} utilized prior quantum adders from \cite{Thapliyal_2016}, which can be, at best, logarithmic in T-depth and therefore, rendering the overall T-depth of the division circuit to be of order $\mathcal{O}(n\log n)$. We ignore the calculation reported in \cite{Thapliyal_2021}, which uses a non-standard way for computing T-depth, resulting in constant-depth adder and linear-depth division.
Interestingly, slow division methods stand out for their cost-effectiveness among all these designs. We made similar observations in this work. More importantly, we showed that a detailed design space exploration can help in identifying highly optimized designs, which has not been undertaken in any prior works.

In this paper, we present the DSE methodology in constructing our quantum non-restoring integer division circuit, where all arithmetic operations are expressed in terms of addition modules. 
The detailed workflow will be provided in the following sections.

\section{Quantum Non-restoring Division}\label{sec: NRBD}

\begin{figure}[ht!]
    \centering
    \includegraphics[width=1.00\linewidth]{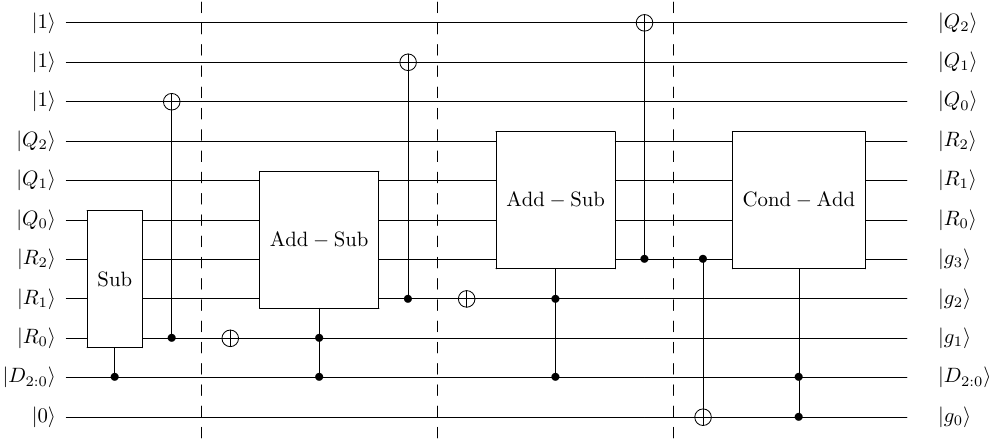}  
    \caption{3-qubits Quantum Non-restoring Divider.\label{fig: Overall}}
\end{figure}

In this section, we describe the construction workflow of our addition-based non-restoring divider, divided into a Quotient Generation Phase and Remainder Restoration Phase. The detailed pseudocode and quantum circuit are presented in Algorithm \ref{Algorithm_NR} and Fig. \ref{fig: Overall}, respectively.

\begin{algorithm}[hbt!]
\caption{Non-restoring Division($Q$,$D$)\label{Algorithm_NR}}
\begin{algorithmic}
\Require{Non-zero unsigned integers $Q$ represents the dividend and $D$ represents the divisor.
\\$Q$ and $D$ are two $n$ bit input values.
\\Returns qubits $Q$ as quotient, and qubits $R$ as remainder.}

\State $R = 0^{n+1}$   \Comment{Where $0^{n+1}$ are $n+1$ zero bits and $R$ and $Q$ are qubit pairs.}
\State $Q = Q_{n-1} Q_{n-2} \cdots Q_{1} Q_{0}$
\State $D = 0 D_{n-1} D_{n-2} \cdots D_{1} D_{0}$
\State\textcolor{red}{\slash$*$ Quotient Generation Phase Start $*$\slash}
\For{$i=0$ to $n$}
    \If{$R_{0} = 0$}
        \State $(R,Q) = 2 * (R,Q)$  \Comment{Left-shift operation}
        \State $R = R - d$          \Comment{Subtraction operation}
    \Else
        \State $(R,Q) = 2 * (R,Q)$
        \State $R = R + d$          \Comment{Addition operation}
    \EndIf
    \If{$R_{0} = 0$}                \Comment{Set $Q_{n-1}$ based on $R_{0}$}
        \State $Q_{n-1} = 1$    
    \Else
        \State $Q_{n-1} = 0$
    \EndIf
\EndFor
\State\textcolor{red}{\slash$*$ Quotient Generation Phase End $*$\slash}
\State\textcolor{blue}{\slash$*$ Remainder Restoration Phase Start $*$\slash}
\If{$R_{0} = 1$}                    \Comment{Check for valid remainder}
    \State $R = R + d$
\EndIf
\State\textcolor{blue}{\slash$*$ Remainder Restoration Phase End $*$\slash}
\\
\textbf{Return} $Q$, $R$
\\\textbf{End}
\end{algorithmic}
\end{algorithm}

\begin{figure}[ht!]
    \centering
    \subfigure[3-qubits Sub Module.\label{Sub}]{\includegraphics[width=0.32\linewidth]{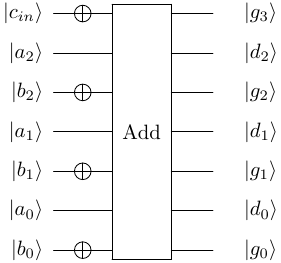}}
    \hfill
    \subfigure[3-qubits Add-Sub Module. \label{AddSub}]{
    \includegraphics[width=0.49\linewidth]{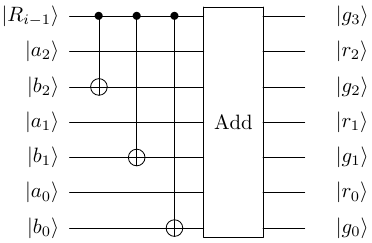}
    }
        \hfill
    \subfigure[3-qubits Cond-Add Module. \label{CondAdd}]{
    \includegraphics[width=0.9\linewidth]{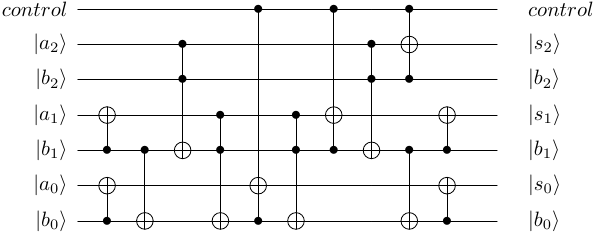}
    }
    \caption{The three essential sub-circuits in the proposed quantum divider.\label{fig:Sub_Circuits}}
\end{figure}

For our design, the essential sub-circuits include the Subtractor, Controlled Adder-Subtractor, and Conditional Adder. For an $n$-qubit division, the adder building blocks and their respective sub-circuits are of $n+1$ qubit-width.
As shown in Fig. \ref{Sub}, the Subtractor sub-circuit effectively transforms quantum adders into subtractors with inputs $\ket{a}$ and $\ket{b}$.
Specifically, the subtraction $\ket{a} - \ket{b} = \ket{a} + \ket{\bar{b}} + 1$, is achieved by flipping the subtrahend bits and setting the carry-in to high. 
In Fig. \ref{AddSub}, we present the implementation of our Controlled Adder-Subtractor sub-circuit. The input bit $\ket{{R}_{i-1}}$, where $i$ represents the current iteration in the Quotient Generation Phase, functions as both the control and carry-in qubit. When $\ket{{R}_{i-1}}$ is equal to $\ket{1}$, it flips the subtrahend qubits for subtraction, performing addition otherwise.
Furthermore,  our Conditional Adder module is illustrated in Fig. \ref{CondAdd} based on previous implementations ~\cite{Thapliyal_2017}. When the control qubit is in the high state, this circuit produces the sum $\ket{s} = \ket{a} + \ket{b}$. Otherwise, $\ket{a}$ and $\ket{b}$ pass through the circuit unchanged.

In the optimization of our workflow, we observe that the first iteration of the Quotient Generation Phase always involves subtraction due to the initialized constant value (0) of qubit \texttt{R} post-initialization. 
Besides, the left-shift operation of the qubit pair (\texttt{R}, \texttt{Q}) can be omitted by appropriately offsetting the sub-circuits of subsequent operations. 
These optimizations result in substantial conservation of quantum resources: (i) reduction in the width of \texttt{R}, (ii) partial execution of the first left-shift operation, and (iii) simplifying the initial arithmetic sub-circuit into a Quantum Subtractor.

As shown below, Equation \ref{formula: TD_Uncomputation}, \ref{formula: TC_Uncomputation}, and \ref{formula: QC_Uncomputation} show the overall cost of the proposed quantum circuit for an $n$-qubit division. This analysis is divided into three steps: Step 1 corresponds to the initial iteration of the Quotient Generation Phase, Step 2 presents the subsequent iterations of the Quotient Generation Phase, and Step 3 shows the Remainder Restoration Phase.
%
\begin{equation}\label{formula: TD_Uncomputation}
\begin{split}
\allowdisplaybreaks
\text{TD}
=&\overbrace{TD_{add}}^{\text{Step 1}}+\overbrace{(TD_{add})\cdot(n-1)}^{\text{Step 2}}+\overbrace{3n+1}^{\text{Step 3}}\\
=&n\cdot TD_{add} +3n+1
\end{split}
\end{equation}
%
\begin{align}\label{formula: TC_Uncomputation}
\begin{split}
\allowdisplaybreaks
\text{TC}
=&\overbrace{TC_{add}}^{\text{Step 1}}+\overbrace{(TC_{add})\cdot(n-1)}^{\text{Step 2}}
+\overbrace{3n+1}^{\text{Step 3}}\\
=&n\cdot TC_{add} +3n+1
\end{split}
\end{align}
%
\begin{equation}\label{formula: QC_Uncomputation}
\begin{split}
\allowdisplaybreaks
\text{QC}
=&\overbrace{3n+2+Anc_{add}}^{\text{Step 1}}+\overbrace{n-1}^{\text{Step 2}}
+\overbrace{1}^{\text{Step 3}}\\
=&4n+2+Anc_{add}\\
\end{split}
\end{equation}
Our design is presented at a high-level model, affording flexibility in the choice of the Quantum Adder. Hence, ${TC}_{add}$ and ${TD}_{add}$ denote the Toffoli Count and Toffoli Depth of the selected adder for an $n+1$ qubit-width sub-module. Similarly, the variable ${Anc}_{add}$ represents the ancilla qubits required by the chosen adder.

\subsection{An Extension Design: Quantum Restoring Divider}

As an extension of our research, we propose an addition-based restoring quantum divider, drawing from the workflow in the previous section. For $n$-qubits division, the corresponding TD, TC, and QC are formulated as $(n\cdot TD_{add} +3n^{2}+n)$, $(n\cdot TC_{add} +3n^{2}+n)$, and $(4n+1+Anc_{add})$, respectively. 
%

As shown in Table \ref{tab:table_Compare_2},
with a carefully selected addition block, our non-restoring design consistently outperforms restoring divider in both TD and TC, with the addition of only one ancilla.
Therefore, we only focus on the non-restoring divider for its superior overall efficiency. 
A detailed description of this extension will be accessible online upon this paper's acceptance.


\section{Results and Discussions }\label{sec: Resu}

\subsection{Impact of Quantum Adders on Quantum Dividers}
\begin{table*}[ht!]
\centering
\renewcommand\arraystretch{1.1} 
\caption{Performance of the proposed $n$-qubits quantum non-restoring divider based on different addition building blocks.\\
The equation for $\omega(n)$ is $\omega(n)=n-\sum_{y=1}^\infty\left \lfloor\frac{n}{2^y}\right \rfloor$ and  $r$ represents the radix, with a range of $2 < r \leq n$.
\label{tab:table_Compare}}
\resizebox{1.0\textwidth}{!}{
\begin{tabular}{|cc|c|c|c|}\hline
    \multicolumn{5}{|c|}{\textbf{Quantum Division Circuit}}\\\hline
    \multirow{2}{*}{\textbf{Constituent Adder Circuit}}&\multirow{2}{*}{\textbf{Year}} &\multicolumn{3}{|c|}{\textbf{\hspace{9 mm}Performance}}\\\cline{3-5}
    & &\textbf{Toffoli Depth} &\textbf{Toffoli Count} &\textbf{Qubit Count} \\\hline

    Variable/Unspecified Adder &$-$ &$n\cdot TD_{add} +3n+1$ &$n\cdot TC_{add} +3n+1$ &$4n+2+Anc_{add}$ \\\hline

    VBE RCA \cite{VBE} &$1995$ &$4n^{2}+5n+1$ &$4n^{2}+5n+1$ &$5n+6$ \\\hline

    Cuccaro RCA \cite{Cuccaro} &$2004$ &$2n^{2}+4n+1$ &$2n^{2}+4n+1$ &$4n+6$ \\\hline

    Draper In-place CLA \cite{Draper} &$2004$ &\makecell{$11n+ n\left \lfloor\log n\right \rfloor+ n\left \lfloor\log (n+1)\right \rfloor$\\$+n \left \lfloor\log \frac{n}3\right \rfloor+n\left \lfloor\log \frac{n+1}3\right \rfloor+1$} &\makecell{$10n^{2}-3n\cdot\omega(n)-3n\cdot\omega(n+1)$\\$-3n\left \lfloor\log n\right \rfloor-3n\left \lfloor\log (n+1)\right \rfloor+6n+1$ } &$6n-\omega(n+1)-\left \lfloor\log (n+1)\right \rfloor+6$ \\\hline

    Takahashi Low-ancilla Adder \cite{Takahashi08} &$2008$ &$30n\log (n+1)+3n+1$ &$28n^{2}+31n+1$ &$4n+\frac{3n+3}{\log (n+1)}+4$ \\\hline

    Takahashi RCA \cite{Takahashi09} &$2009$ &$2n^{2}+4n+1$ &$2n^{2}+4n+1$ &\textcolor{red}{$4n+5$} \\\hline

    Takahashi Combination \cite{Takahashi09} &$2009$ &$18n\log (n+1)+3n+1$ &$7n^{2}+10n+1$ &$4n+\frac{3n+3}{\log (n+1)}+4$ \\\hline

    Wang RCA
    \cite{paper4_2016} &$2016$ &$n^{2}+4n+1$ &\textcolor{red}{$n^{2}+4n+1$ }&$5n+6$ \\\hline

    Gidney RCA
    \cite{Gidney2018halvingcostof} &$2018$ &$n^{2}+4n+1$ &$2n^{2}+3n+1$ &$5n+4$ \\\hline

    Gayathri RCA
    \cite{paper3_2021} &$2021$ &$n^{2}+4n+1$ &\textcolor{red}{$n^{2}+4n+1$ }&$5n+6$ \\\hline

    Higher Radix Adder \cite{wang2023higher} &$2023$ &\begin{tabular}[c]{@{}c@{}}$4 n\log (n+1)+3r\cdot n-2n\log r$\\$ -2n\log 3r+2n\log (r-2)+5n+1$\end{tabular} &\begin{tabular}[c]{@{}c@{}}$8n^{2}-\frac{n(n+1)}{r}-n^{2}\pmod  r$\\$-3n\cdot \omega(\frac{n+1}{r})-3n\log (n+1) +3n\log r +8n+1$\end{tabular} &\begin{tabular}[c]{@{}c@{}}$6n-\log (n+1)+\frac{n+1}{r}$\\$ -\omega(\frac{n+1}{r})+\log r+5$\end{tabular} \\\hline
 
    Quantum Ling Adder \cite{wang2023reducing} &$2023$ &\textcolor{red}{$12n+2n\lfloor\log\frac{n+1}{2} \rfloor+2n\lfloor\log\frac{n+1}{6} \rfloor+1$} &$13n^{2} -6n\cdot \omega(\frac{n+1}{2})-6n\lfloor \log{\frac{n+1}{2}} \rfloor+2n+1$ &$14n-6\omega(\frac{n+1}{2})-6\lfloor \log{\frac{n+1}{2}} \rfloor+4$ \\\hline

\end{tabular}}
\end{table*}

\begin{figure}[ht!]
    \centering
    \subfigure[Low Toffoli Depth Non-restoring Divider. \label{fig:TD}]{\includegraphics[width=0.8\linewidth]{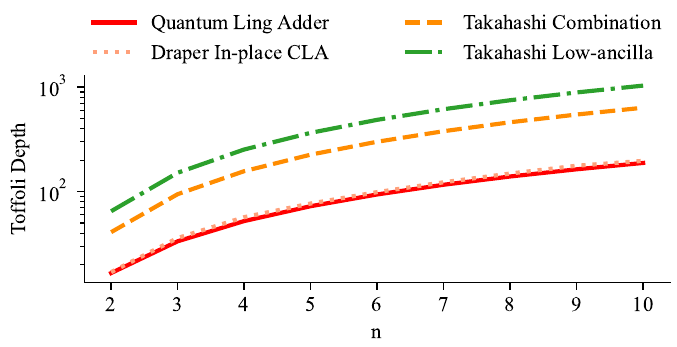}
    }
    \hfill
    \subfigure[Low Toffoli Count Non-restoring Divider. \label{fig:TC}]{
    \includegraphics[width=0.8\linewidth]{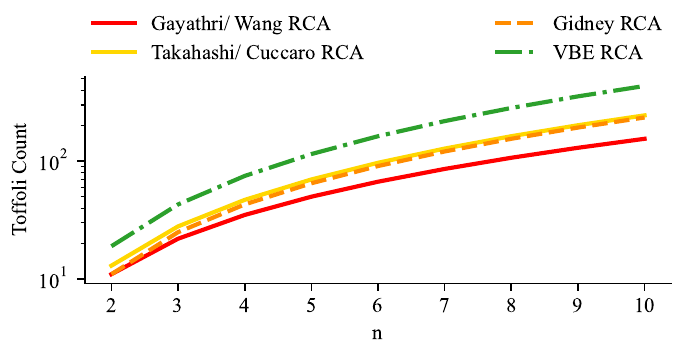}
    }
        \hfill
    \subfigure[Low Qubit Count Non-restoring Divider. \label{fig:QC}]{
    \includegraphics[width=0.8\linewidth]{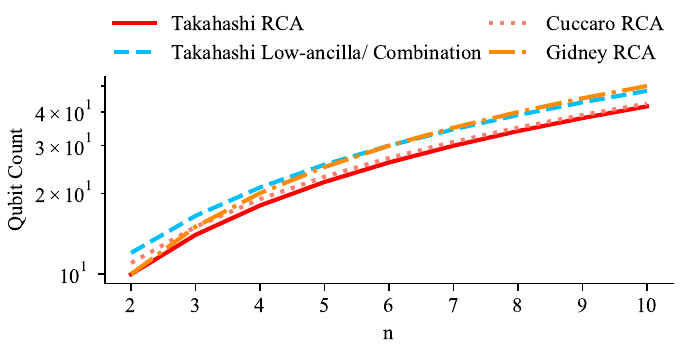}
    }
    \caption{The proposed quantum non-restoring dividers based on different adders.
    Here, $n$ denotes the qubit-width of the division. Given the identical performance in Toffoli Count between Gayathri RCA and Wang RCA, we use a simplified notation "Gayathri/Wang RCA" to represent their performance. The similar notations are also applied for "Takahashi/Cuccaro RCA" and "Takahashi Low-ancilla /Combination".\label{fig:Comparison}}
\end{figure}

As shown in Table \ref{tab:table_Compare}, the selection of the adder building block has a substantial impact on the performance of our divider.
Here, we systematically evaluated various building blocks based on TC, TD and QC to identify the most suitable choices for different scenarios.
\begin{itemize}
    \item \textbf{Toffoli Depth.} Toffoli Depth significantly influences the execution time of the quantum circuit. 
    As shown in Table \ref{tab:table_Compare}, CLA adders usually achieve low Toffoli Depth by using the parallel prefix tree approach to maximize parallel computation sub-blocks and reduce calculation depth. 
    For example, using Draper In-place CLA as the basic building block achieves superior Toffoli Depth compared to other quantum RCAs and Combination Adders such as Takahashi RCA and Takahashi Combination Adder.
    Significantly, the Quantum Ling Adder-based non-restoring divider achieves a substantial reduction in Toffoli Depth complexity from $O(n\log(n))$ to $O(n\log(\frac{n}{2}))$.
    
    \item \textbf{Toffoli Count.} When the primary target is to minimize the Toffoli Count, RCA-based addition building blocks offer a great solution. 
    As shown in Fig \ref{fig:TC} and Table \ref{tab:table_Compare}, these RCA-based building blocks are characterized by their simplicity and low Toffoli Count, making them an attractive choice for Toffoli Count-sensitive scenarios.
    Among all the building blocks, it is noteworthy that the Gayathri RCA and Wang RCA achieve the lowest costs in terms of Toffoli Count by using only $n$ Toffoli gates to complete $n$-qubits addition.
    
    \item \textbf{Qubit Count.} In quantum computing, evaluating the Qubit Count is crucial for efficient quantum resource management. Although CLA-based designs achieve low Toffoli Depth, they usually require more ancilla qubits.
    As illustrated in Figure \ref{fig:QC}, RCA-based building blocks clearly demonstrate their advantages in achieving competitive Qubit Count.
    Apart from the RCA-based building blocks, an appealing alternative is Takahashi's Combination Adder \cite{Takahashi09}, which strategically integrates components from both RCA and CLA, resulting in a low Qubit Count that scales linearly with the division bitwidth $n$.
    Remarkably, Takahashi's Low-ancilla Adder \cite{Takahashi08} has the same Qubit Count as Takahashi's Combination Adder \cite{Takahashi09}, making it an equally appealing choice.
    Therefore, RCA-based adders, \cite{Takahashi08} and \cite{Takahashi09} are highly recommended as the preferred building blocks in QC cost-dominated situations.
\end{itemize}

Overall, as illustrated in Figures \ref{fig:TD}, \ref{fig:TC}, and \ref{fig:QC}, the most effective addition building blocks for minimizing Toffoli Depth, Toffoli Count, and Qubit Count are identified as Quantum Ling adder, Gayathri/Wang RCA, and Takahashi RCA, respectively.

\subsection{Comparison with Existing Quantum Dividers}
Following a comprehensive design space exploration to assess the impact of quantum adders on our design, we constructed and compared the proposed restoring and non-restoring dividers with existing work.
The Quantum Ling Adder which has minimal Toffoli Depth achieves $94.06\%$ in Toffoli Depth savings, and the Takahashi Combination Adder, which offers balanced results across all metrics sees $91.98\%$ and $99.37\%$ savings in Toffoli Count and Qubit Count respectively.
%
\begin{table}[ht!]
\centering
\caption{Comparison of Proposed 32-qubits Quantum Dividers with Existing Designs.
In this table, "Takahashi C" refers to the Takahashi Combination Adder.
Percentage Improvement is calculated w.r.t the Quantum Newton Raphson divider \cite{Gayathri_2022}
\label{tab:table_Compare_2}}
\resizebox{0.48\textwidth}{!}{
\begin{tabular}{|c|c|c|c|}
\hline \textbf{Divider} & \textbf{Toffoli Depth} & \textbf{Toffoli Count} & \textbf{Qubit Count} \\\hline
 
 Goldschmidt \cite{Gayathri_2021} &$17,850$ &$117,187$ &$30,008$ \\\hline
 
 Newton Raphson \cite{Gayathri_2022} &$13,506$ &$93,376$ &$23,996$ \\\hline

 \rowcolor{Celadon!36} Restoring$+$Ling              &$3,809\:(71.80\%)$  &$15,224\:(83.70\%)$  &$415\:(98.27\%)$ \\\hline
 
 \rowcolor{Celadon!36} Restoring$+$ Takahashi C      &$6,010\:(55.50\%)$  &$10,496\:(88.76\%)$  &\textcolor{red}{$151\:(99.37\%)$} \\\hline
 
 \rowcolor{Celadon!36} Non-restoring$+$Ling          &\textcolor{red}{$802\:(94.06\%)$}    &$12,217\:(86.92\%)$   &$416\:(98.27\%)$ \\\hline 
 
 \rowcolor{Celadon!36} Non-restoring$+$ Takahashi C  &$3,003\:(77.77\%)$  &\textcolor{red}{$7,489\:(91.98\%)$}   &$152\:(99.37\%)$ \\\hline
 
\end{tabular}}
\end{table}

The specific data for $32$-qubits division is shown in Table \ref{tab:table_Compare_2}.
Obviously, our design outperform all the prior designs in terms of TD, TC and QC, owing to the absence of resource-intensive conditional statements and operations in the proposed workflows.


\section{Conclusion}\label{sec: Conc}
In conclusion, this paper explored the efficiency of quantum non-restoring dividers, focusing on the impact of integrating different quantum adders. This required an in-depth study of the division algorithm, expressing it in terms of addition circuits.
We observed that when minimizing the Toffoli Depth is crucial, a divider based on quantum CLA adders offers Toffoli Depth reductions of up to $94.06\%$. When minimizing Toffoli Count is essential, the quantum RCAs are a more apt choice, achieving reductions up to $91.98\%$. Lastly, both RCA-based adders and the Takahashi Combination Adder offer optimal solutions for prioritizing Qubit Count at up to $99.37\%$ improvement.
For ease of future benchmarking, the Qiskit code for the proposed divider will be released upon this paper's acceptance.

In future, there are various promising directions in this field.
One of the most crucial directions is optimizing specific quantum divider designs for diverse applications based on our work.
Moreover, assessing the performance and scalability of these proposed dividers in different quantum algorithms holds significant importance. 
Furthermore, conducting extensive relevant experiments on quantum computers is essential for transitioning from simulations to practical implementations.

\section*{Acknowledgement}
This research is supported by the National Research Foundation, Singapore under its Quantum Engineering Programme Initiative. Any opinions, findings and conclusions or recommendations expressed in this material are those of the authors and do not reflect the views of National Research Foundation, Singapore.

\bibliographystyle{IEEEtran}
\bibliography{main}
\vspace{12pt}

\end{document}